\documentclass[twocolumn,showpacs,preprintnumbers,amsmath,amssymb]{revtex4}
\usepackage{graphicx}
\usepackage{dcolumn}
\usepackage{bm}
\usepackage{subfigure}

\begin{document}


\title{Gauge-invariant construction of quantum cosmology}

\author{Fumitoshi Amemiya}
 \email{famemiya@rk.phys.keio.ac.jp}
\author{Tatsuhiko Koike}
 \email{koike@phys.keio.ac.jp}
\affiliation{Department of Physics, Keio University, 3-14-1 Hiyoshi, Kohoku-ku, 223-8522 Yokohama, Japan }

\date{\today}

\begin{abstract}
We present and analyze a gauge-invariant quantum theory of the Friedmann-Robertson-Walker universe with dust.
We construct the reduced phase space spanned by gauge-invariant quantities by using the so-called relational formalism at the classical level.
The reduced phase space thereby obtained can be quantized 
in the same manner as an ordinary mechanical system. 
We carry out the quantization and obtain the Schr\"{o}dinger equation. 
This quantization procedure realizes a possible resolution to the problem of time and observables in canonical quantum gravity.
We analyze the classical initial singularity of the theory by 
evolving a wave packet backward in time and evaluating the expectation value of the scale factor. 
It is shown that the initial singularity of the Universe is avoided by the quantum gravitational effects. 
\end{abstract}

\maketitle
\section{Introduction \label{sec1}}
On large scales, observational data show that the Universe is well described by general relativity (GR).
Since the Universe is expanding at present, it becomes smaller as it evolve backward in time and at last we encounter the initial singularity.
However, it is thought that quantum gravitational effects become dominant at the Planck scale, where the Universe itself must be treated as a quantum object.
One thus needs a quantum theory of the Universe in order to understand what happens near the initial singularity.
Symmetry reduced models, e.g., the isotropic and homogeneous model~\cite{Hawking1, Hawking2, Vilenkin1, Vilenkin2}, have been canonically quantized by the Dirac quantization method, which leads to the Wheeler-DeWitt equation~\cite{WDW1,WDW2}. 
Unfortunately, this method causes the ``problem of time'' (see, e.g.,~\cite{pot} and references therein), that is, time evolution is lost in the following two senses.
The first is the dynamics of the wave function $\Psi$.
The canonical formulation of GR leads to a constrained system and a Hamiltonian is of the form $H=\Sigma_{i}N^i C_i$, where $C_i=0$ are first-class constraints and $N^i$ are arbitrary functions.
Since the classical constraints are translated into the quantum constraints $\hat{C}_i \Psi =0$ by the Dirac method, the Schr\"{o}dinger equation becomes trivial for the physical state: $i\hbar \frac{\partial \Psi}{\partial t}=\hat{H}\Psi=0$.
Therefore, the dynamics of the wave function is lost in the quantum framework.
The second is concerned with observables.
In ordinary gauge theories, observables are gauge-invariant quantities. 
Since a gauge-invariant quantity $O$ is defined as having vanishing Poisson brackets with all constraint functions $\{C_i,O\}=0$, the quantity $O$ becomes a constant of motion by the Hamilton equation $\dot{O}=\{H,O\}=0$.
Thus, the dynamics of observables is lost in both classical and quantum theories if one restricts observables to gauge-invariant quantities.  
In fact, the problem of what should be interpreted as observables in classical GR has long been discussed by many researchers, see for instance~\cite{Einstein, Bergmann, DeWitt, Stachel, Earman, Rovelli, Lusanna}.

A possible way to overcome these problems is that one finds gauge-invariant quantities and a method to extract their physical evolution at the classical level, and then constructs a quantum theory based on them.
The idea for constructing gauge-invariant quantities which has been stressed in~\cite{Bergmann, Rovelli} is that the relation between dynamical variables is gauge invariant even if they are gauge variant, respectively. 
As the realization of the idea, a formal expression of gauge-invariant quantities in constrained systems was recently presented~\cite{BD1,reduced quantization}.
This method is often called the relational formalism.
Although the application of the formalism to full GR is hard in general, the situation
is dramatically simplified if the constraints can be written in the so-called deparametrized
form (see, e.g.,~\cite{deparametrization}).  
If one applies the relational formalism to a deparametrized theory~\cite{Phantom field, Phantom dust}, one can construct the reduced phase space coordinatized
by gauge-invariant quantities and obtain a physical Hamiltonian which generates
the time evolution thereof. 
Then, one can quantize the reduced system in the
same manner as in elementary quantum mechanics and obtain the Schr\"{o}dinger equation because there are no constraints in the reduced phase space. 

In the present work, we construct and analyze a gauge-invariant quantum theory of the Friedmann-Robertson-Walker (FRW) universe without the problem of time.
We consider the case when the matter involves dynamical dust coupled to gravity, where the dust action is introduced by Brown and Kucha\v{r}~\cite{BK}.
The advantage of the dust is that one can deparametrize the Hamiltonian constraint and extract a natural time variable which corresponds to the cosmological proper time when one solves equations of motion.
Therefore, we can apply the relational formalism to construct the reduced phase space of the FRW universe with dust, and then quantize the reduced system without dealing with the constraint.  
In order to investigate what happens near the classical initial singularity, we find and analyze the solutions of the constructed quantum theory of the Universe. 
We first construct a wave packet which allows semiclassical interpretation.  
Then, we evolve it backward in time by the Schr\"{o}dinger equation and evaluate the expectation value of the scale factor. 
As a consequence, it is shown that the initial singularity is replaced by a big bounce by the quantum gravitational effectsD
 
The organization of the paper is as follows.
In Sec. \ref{sec3} we review the relational formalism for deparametrized
theories and present the classical theory of the FRW universe with the dust in Sec. \ref{sec4}. 
Then, we quantize the obtained system in Sec. \ref{sec5}, where we divide the section into three parts by the operator ordering of the Hamiltonian.
We analyze the dynamics of the Universe in Sec. \ref{sec6}. 
Sec. \ref{sec7} is for the conclusion. 

In this paper, we adopt the following unit for the speed of light: $c=1$.

\section{Relational formalism for deparametrized theories \label{sec3}}
In this section, we summarize the relational formalism in the case
where the constraint can be written in the deparametrized form. See~\cite{BD1,
  reduced quantization} for general cases and details. 
From now on, we assume that the system has only one constraint for simplicity.

\begin{figure}
\includegraphics[width=55mm]{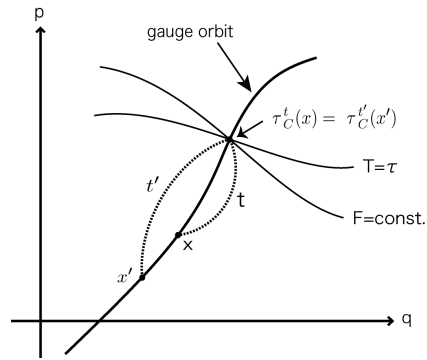}
\caption{\label{alpha}The action of the map $\alpha_C^t$ on phase space points and the gauge-invariance of $O_{F}^{\tau}(x)$. }
\end{figure}

The key observation of the relational formalism to define gauge-invariant
quantities is as follows.  
Take two functions $F$ and $T$ on the phase space.
Then, the value of $F$ at $T=\tau$ is gauge invariant even if $F$ and $T$
themselves are gauge variant. 
That is, one can interpret one of the functions $T$ as a clock and consider the relation between $T$ and other variables as time evolution. 
If we denote a phase space point by $x=(q^a,p_a)$, the mathematical definition
of the gauge-invariant quantity $O_F^\tau(x)$ as a phase space function is given
by  
\begin{align}
O_{F}^\tau(x) := \alpha_C^t(F)(x) | _{\alpha_C^t(T)(x)=\tau}.\label{gi}
\end{align}
Here, $\alpha_{C}^{t}$ denotes the action of the gauge transformation generated
by $C$,  and $t$ is a parameter along the gauge orbits.
Thus, $\alpha_C^t(x)$ is the gauge flow generated by $C$ starting from $x$. 
The action of $\alpha_{C}^{t}$ on a function is given by
$\alpha_{C}^{t}(F)(x)=F\left(\alpha_{C}^{t}(x)\right)$,
which is written as a series
$\alpha_{C}^{t}(F)(x) = \sum_{n=0}^{\infty}\frac{t^n}{n!}\{C,F\}_{(n)}(x)$,
where $\{C,F\}_{(0)}:=F$ and $\{C,F\}_{(n+1)}:=\left\{C,\{C,F\}_{(n)} \right\}$. 
The definition (\ref{gi}) gives a manifestly gauge-invariant quantity because
$O_F^\tau(x)$ is constant on each gauge orbit.
Indeed, if $x$ and $x^{\prime}$ are on the same gauge orbit, there is $t^{\prime}$ such that $\alpha_{C}^{t}(x)=\alpha_{C}^{t^{\prime}}(x^{\prime})$ and 
$O_F^{\tau}(x)=F\left(\alpha_C^t(x)\right)\bigr|_{T\left(\alpha_C^t(x)\right)=\tau}\nonumber=F\left(\alpha_C^{t^{\prime}}(x^{\prime})\right)\Bigr|_{T\left(\alpha_C^{t^{\prime}}(x^{\prime})\right)=\tau}=O_F^{\tau}(x^{\prime})$ (see Fig.\ref{alpha}).

A constraint equation $C=0$ is said to be of deparametrized form if it is written as 
\begin{align}
C=P+h(q^a,p_a)=0
\end{align}
with some phase space coordinates $\{q^a,T;p_a,P\}$. 
In the deparametrized theories, the reduced phase space is spanned by the gauge-invariant quantities $\left(O_{q^a}^\tau(x),O_{p_a}^\tau(x)\right)$ associated with $q^a$ and $p_a$ with the simple symplectic structure
\begin{align}
\left\{O_{q^a}^\tau(x),O_{p_b}^\tau(x)\right\}=\delta^a_b.\label{poisson}
\end{align}
The physical Hamiltonian $H$ is obtained by replacing $q^a$ and $p_a$ in $h(q^a,p_a)$ with $O_{q^a}^\tau(x)$ and $O_{p_a}^\tau(x)$,
\begin{align}
H\left(O_{q^a}^\tau(x),O_{p_a}^\tau(x)\right):=h\left(O_{q^a}^\tau(x),O_{p_a}^\tau(x)\right),\label{H}
\end{align}
which generates the time evolution of the gauge-invariant quantities:
\begin{align}
\frac{\partial O_{F}^\tau(x)}{\partial \tau}= \left\{H,O_{F}^\tau(x)\right\}.\label{hamilton}
\end{align}
As we have seen, one can construct the gauge-invariant quantities and extract their physical evolution in the deparametrized theories by using the relational formalism.

\section{Friedmann-Robertson-Walker universe with dust \label{sec4}}
In this section, we shall formulate the FRW universe with dust.
As was mentioned above, the advantage of the dust introduced by Brown and
Kucha\v{r}~\cite{BK} is that one can deparametrize the system. 

The action of the dust is given by
\begin{align}
S_{\textrm{dust}}=-\frac{1}{2}\int_{M}d^4x \sqrt{-g}\rho
(g^{\mu\nu}U_{\mu}U_{\nu}+1), 
\end{align}
where the one-form $U_\mu$ is defined by $U_\mu=-(dT)_\mu+W_j(dZ^j)_\mu$ ($j=1,2,3$), $\rho$ is the rest mass density of the dust and $g$ is the determinant of the metric tensor $g_{\mu\nu}$ on the spacetime manifold $M$.
The action is written by the variables $g_{\mu\nu}$, $\rho$, $T$, $Z^j$ and $W_j$.
When the equations of motion hold, 
$Z^j$ labels the flow lines of the dust particle and 
$T$ is the proper time along them. 
By using the Dirac algorithm for constrained systems, 
one can obtain the canonical form of the dust action
\begin{align}
S_{\textrm{dust}} = \int dt \int_{\Sigma} d^3x \big( &P\dot{T} + P_j \dot{Z}^j\nonumber\\
&- NC_{\textrm{dust}} -N^aD_a^{\textrm{dust}} \big), 
\end{align}
with the dust degrees of freedom being $T$ and $Z^j$ and their conjugate
momenta $P$ and $P_j$, respectively, 
where 
\begin{align}
&C_{\textrm{dust}}=\sqrt{ P^2+q^{ab}D_a^{\textrm{dust}}D_b^{\textrm{dust} }},\\
&D_a^{\textrm{dust}}=PT_{,a}+P_jZ^j_{,a}. 
\end{align}
Here, $N$ and $N^a$, respectively, are the lapse function and the shift vector and $q_{ab}$ is the induced metric on a three-space $\Sigma$. 
It is worth noting that $P$ is written by using the original variables as 
$P=-\rho\sqrt{\det\left(q_{ab} \right)} U^{\mu}n_{\mu}$,
where $n^{\mu}$ is the future-pointing unit normal to $\Sigma$. 
It is also shown that $U^{\mu}$ is a future-pointing unit timelike vector on shell, so that $U^{\mu}n_{\mu}<0$ and the signs of $P$ and $\rho$ are the same.
Recently, several applications of the relational formalism to systems with the dust have appeared~\cite{Phantom dust, manifest1, manifest2}, where the dust has negative energy in order for the physical Hamiltonian to become positive definite.     
In our work, we choose the signs of $P$ and $\rho$ as positive in order to interpret the dust as a standard matter.
Although this choice leads to a negative Hamiltonian, it is not surprising because the contribution of the gravitational fields to the Hamiltonian is originally negative.

In the case of the FRW universe, the dust action is cast into the form 
\begin{align}
S_{\textrm{dust}}=\int dt \int_{\Sigma}d^3x \left( P\dot{T}-NC_{\textrm{dust}}\right),
\end{align}
where
\begin{align}
C_{\textrm{dust}}=P.
\end{align}
Here, $T$ corresponds to the cosmological proper time when one solves the equations of motion.
Therefore, it can be interpreted as a natural time.
This will be used as a clock in the discussion below.

On the other hand, the action of gravity in the FRW spacetime is given by
\begin{align}
S_{\textrm{grav}} = \int dt \int_{\Sigma} d^3x \left( p_a \dot{a}  -NC_{\textrm{grav}} \right) ,
\end{align}
where 
\begin{align}
C_{\textrm{grav}} = - \kappa\frac{ p^2_a}{12a} + \frac{\Lambda a^3}{\kappa} - \frac{3ka}{\kappa}+\frac{R}{a}.
\end{align}
Here, $a$ is the scale factor, $p_a$ is its conjugate momentum defined as $p_a:=-\frac{6a\dot{a}}{\kappa  N}$, $\Lambda$ is the cosmological constant, $R=\rho_{\textrm{rad}}a^4$ is a constant associated with the kinematically incorporated radiation with energy density $\rho_{\textrm{rad}}$ and the constant $k=\pm 1,0$ determines the curvature of the three-dimensional space.
For simplicity, we shall only consider the case of the flat universe, $k=0$. The metric is written in Cartesian coordinates as 
\begin{align}
ds^2=-dt^2+a^2(t)(dx^2+dy^2+dz^2),
\end{align}
where we assume that the scale factor has a dimension of length and the coordinates are dimensionless.

The total action of the FRW universe with the dust takes the form
\begin{align}
S_{\textrm{tot}}&=S_{\textrm{grav}}+S_{\textrm{dust}}\nonumber\\
&=\int dt \int_{\Sigma} d^3x \left( p_a \dot{a} + P\dot{T} -NC_{\textrm{tot}} \right),\\
C_{\textrm{tot}}&=C_{\textrm{grav}}+C_{\textrm{dust}}=C_{\textrm{grav}}+P.
\end{align}
Here, the constraint is of the almost deparametrized form except for the existence of the 
three-space integral. 
In general, since the volume of the three-space will diverge in the flat case, 
we must somehow get rid of this divergence in order to deparametrize the system and 
consider the canonical quantization.

We shall avoid the divergence and deparametrize the system by considering a
compact universe. 
For simplicity, we only consider the case of three-dimensional torus, where we
take a cube of coordinate range 
$0\le x,y,z\le L$,
and identify the opposite faces.
Then, the scale factor multiplied by $L$ is the length of an edge of the cube, which, in fact, represents the physical size of the Universe.
If we denote the coordinate volume as $V := \int_{\Sigma}d^3x = L^3$,
the change of the variables
$a^{\prime} := V^{\frac{1}{3}}a=aL$, $p_a^{\prime} := V^{\frac{2}{3}}p_a$ and $P^{\prime} := VP$
absorbs the volume factor in the total action:
\begin{align}
&S_{\textrm{tot}} = \int dt \left[ p^{\prime}_a \dot{a}^{\prime} + P^{\prime}\dot{T}
-NC^{\prime}_{\textrm{tot}} \right],\label{action}\\
&C^{\prime}_{\textrm{tot}}= P^{\prime}+ h(a^{\prime},p_a^{\prime}) ,\\
&h(a^{\prime},p_a^{\prime})=-\kappa  \frac{p^{\prime 2}_a}{12a^{\prime}} + \frac{\Lambda a^{\prime 3}}{\kappa}+\frac{R}{a^{\prime}}.
\end{align}
Here, $a^{\prime}$ is the physical length of an edge of the Universe
(three-torus) and thus $V_{\textrm{phys}}:=a^{\prime 3}=a^3 L^3$ is the physical
volume of the universe. 
So far, we have finished deparametrizing the system of the FRW universe with the dust.

Let us apply the relational formalism to the system and 
obtain the reduced phase space coordinatized by gauge-invariant quantities. 
In the present case, $T$ becomes the clock, that is, the evolution of all
gauge-invariant quantities are measured by a relative relation with respect to
$T$. 
As explained in Sec.\ref{sec3}, the reduced phase space is coordinatized by
\begin{align}
A(\tau):=O_{a^{\prime}}^\tau(x),\ P_{A}(\tau):=O_{p_a^{\prime}}^\tau(x).
\end{align}
Recall that, the meaning of $O_{a^{\prime}}^\tau(x)$ is the value of $a^{\prime}$ at $T=\tau$.
The symplectic structure is written as
\begin{align}
\left\{A(\tau),P_{A}(\tau)\right\}=1.\label{pb} 
\end{align}
As can be seen from (\ref{H}), the Hamiltonian which generates the time
evolution of the gauge-invariant quantities is obtained by replacing $a^{\prime}$
and $p_a^{\prime}$ in $h(a^{\prime},p_a^{\prime})$ with $A(\tau)$ and
$P_A(\tau)$: 
\begin{align}
H=h(A,P_A)
=-\kappa  \frac{P^{2}_A}{12A} + \frac{\Lambda A^{3}}{\kappa}+\frac{R}{A}.\label{hamilt}
\end{align}

We have obtained the reduced phase space of the FRW universe with the dust. 
The system now is characterized by (\ref{pb}) and (\ref{hamilt}) with no constraints, so that we can quantize the system in the same procedure as in the ordinary canonical quantization.
Hereafter, we call the gauge-invariant quantity $A(\tau)$ the scale factor.

\section{Quantization \label{sec5}}
Let us now quantize the system obtained in the previous section. 
Our classical system is one-dimensional with the symplectic structure 
(\ref{pb}). 
In the ordinary procedure of the canonical quantization, the Poisson bracket is
replaced by the commutation relation among the operators corresponding to
canonical variables: 
\begin{align}
[\hat{A},\hat{P}_{A}]=i\hbar. 
\end{align}
We take the ordinary Schr\"{o}dinger representation of the canonical commutation
relation, that is, $\hat{A}$ acts on the state $\Psi(A)$ in the Hilbert space defined below by multiplication, 
and $\hat{P}_{A}$ acts on $\Psi(A)$ by differentiation: 
\begin{align}
\hat{A}\Psi(A)= A\Psi(A),\ \hat{P}_A\Psi(A)= -i\hbar\frac{\partial}{\partial A}\Psi (A).
\end{align}
In general, there are many ambiguities in the choices of canonical variables, the Hilbert space and the operator ordering. 
As for the first one, we have chosen the most natural phase space variables, the scale factor
and its conjugate momentum. 
Although this choice leads to negative values of the scale factor by von Neumann's theorem, one can consider the restriction of the range of the scale factor $A\ge 0$ in the quantum system. 
Under the restriction, $\hat{P}_A$ fails to be self-adjoint in general.
However, one can ensure the self-adjointness of the Hamiltonian by imposing boundary conditions on wave functions.
Thus, we first define the operator ordering of the Hamiltonian (\ref{hamilt}) and the measure on the space of wave functions, and then derive the boundary conditions on wave functions.
We consider the three cases where $\frac{P^2_{A}}{A}$ is promoted to (a) $\hat{P}_A\frac{1}{\hat{A}}\hat{P}_A$, (b) $\frac{1}{3}\left( \hat{P}_A^2\frac{1}{\hat{A}}+\hat{P}_A\frac{1}{\hat{A}}\hat{P}_A+\frac{1}{\hat{A}}\hat{P}^2_A \right)$ and (c) $\frac{1}{\hat{A}}\hat{P}^2_A$.

\subsection{Operator ordering (a) $\frac{P^2_{A}}{A} \to
  \hat{P}_A\frac{1}{\hat{A}}\hat{P}_A$}
\label{ordering-a}
In the case (a), the Hamiltonian operator is given by
\begin{align}
\hat{H}=-  \frac{\kappa}{12} \hat{P}_A\frac{1}{\hat{A}}\hat{P}_A+ \frac{\Lambda \hat{A}^{3}}{\kappa}+\frac{R}{\hat{A}},\label{ham1}
\end{align}
and the Schr\"{o}dinger equation $i\hbar \frac{\partial \Psi}{\partial \tau} =\hat{H}\Psi
$ takes the form
\begin{align}
i\hbar \frac{\partial \Psi}{\partial \tau} = \frac{\kappa \hbar^2}{12}\left( \frac{1}{A}\frac{\partial^2 \Psi}{\partial A^2} -\frac{1}{A^2}\frac{\partial \Psi}{\partial A} \right) + \left(\frac{\Lambda A^3}{\kappa}+\frac{R}{A}\right)\Psi.\label{sch1}
\end{align}
We define the Hilbert space as $\mathcal{H}=L^2(\mathbb{R}_{+},dA)$ where $\mathbb{R}_{+}$ represents the set of non-negative real numbers.
With this inner product, $\hat{P}_A$ is Hermitian and in fact symmetric, but not self-adjoint.
The condition for the Hamiltonian to become Hermitian $\langle \Psi_1 | \hat{H} | \Psi_2 \rangle = \langle \Psi_1 | \hat{H}^{\dagger} |\Psi_2 \rangle$ is satisfied when
\begin{align}
\int^{\infty}_{0}dA& \left( \Psi_1^{\ast}\frac{1}{A}\frac{d^{2}\Psi_2}{d A^2} -\Psi^{\ast}_1\frac{1}{A^2}\frac{d\Psi_2}{dA} \right) \nonumber\\
&=\int^{\infty}_{0}dA \left( \frac{1}{A}\frac{d^{2}\Psi_1^{\ast}}{d A^2}\Psi_2 - \frac{1}{A^2}\frac{d\Psi_1^{\ast}}{dA}\Psi_2\right).\label{condition}
\end{align}
Integrating the first terms on the both sides by parts, we obtain the equation
\begin{align}
\Psi_1^{\ast}\frac{1}{A}\frac{d\Psi_2}{d A}\biggr|_0^{\infty} =\frac{d\Psi_1^{\ast}}{d A}\frac{1}{A}\Psi_2\biggr|_0^{\infty}. \label{her1}
\end{align}
If we assume the wave functions vanish at infinity, Eq. (\ref{her1}) yields $\Psi_1^{\ast}\frac{1}{A}\frac{d\Psi_2}{d A}\bigr|_{A=0} = \frac{d\Psi_1^{\ast}}{d A}\frac{1}{A}\Psi_2 \bigr|_{A=0}$ and the relation holds if each of the wave functions $\Psi_1$ and $\Psi_2$ satisfies the condition
\begin{align}
\frac{1}{\sqrt{A}}\left(\Psi-\gamma\frac{d \Psi}{dA}\right)\biggr|_{A=0}=0, \label{BC}
\end{align}
where $\gamma$ is a real number.
The Hamiltonian is indeed self-adjoint if its domain is restricted to wave functions which satisfy the boundary condition (\ref{BC}).

As we have seen, the Schr\"{o}dinger equation and the boundary condition to ensure the self-adjointness of the Hamiltonian in the Hilbert space $L^2(\mathbb{R}_{+},dA)$ are derived.
In Sec.V, we shall use the simplest boundary condition corresponding to $\gamma=0$:
\begin{align}
\frac{\Psi (A)}{\sqrt{A}}\biggr|_{A=0}=0.\label{bc2}
\end{align}

For later use, we present the dimensionless form of the Schr\"{o}dinger equation.
Since the characteristic length and time scales here are the Planck length $l_P=\sqrt{\frac{\hbar G}{c^3}}$ and the Planck time $t_P=\sqrt{\frac{\hbar G}{c^5}}$, we have the following dimensionless equation by using them:  
\begin{align}
i\frac{\partial \Psi}{\partial \eta} = \frac{1}{x}\frac{\partial^2 \Psi}{\partial x^2}-\frac{1}{x^2}\frac{\partial\Psi}{\partial x}+\left(\lambda x^3 + \frac{r}{x}\right)\Psi,\label{Schrodinger}
\end{align} 
where $x:=\left(\frac{3}{2\pi}\right)^{\frac{1}{3}}\frac{A}{l_P}$, $\eta:=\frac{\tau}{t_P}$, $\lambda:=\frac{\hbar G}{12 }\Lambda$ and $r:=\left(\frac{3}{2\pi}\right)^{\frac{1}{3}}\frac{R}{\hbar }$.

\subsection{Operator ordering (b) $\frac{P^2_{A}}{A} \to \frac{1}{3}\left( \hat{P}_A^2\frac{1}{\hat{A}}+\hat{P}_A\frac{1}{\hat{A}}\hat{P}_A+\frac{1}{\hat{A}}\hat{P}^2_A \right)$}
In the case (b), the Hamiltonian operator is given by
\begin{align}
\hat{H}=-  \frac{\kappa}{36}\left( \hat{P}_A^2\frac{1}{\hat{A}}+\hat{P}_A\frac{1}{\hat{A}}\hat{P}_A+\frac{1}{\hat{A}}\hat{P}^2_A \right)+ \frac{\Lambda \hat{A}^{3}}{\kappa}+\frac{R}{\hat{A}},\label{ham2}
\end{align}
and the Schr\"{o}dinger equation is written in the form
\begin{align}
i\hbar \frac{\partial \Psi}{\partial \tau} = \frac{\kappa \hbar^2}{12}\left( \frac{1}{A}\frac{\partial^2 \Psi}{\partial A^2} -\frac{1}{A^2}\frac{\partial \Psi}{\partial A} +\frac{2}{3A^3}\Psi \right)\nonumber\\
+ \left(\frac{\Lambda A^3}{\kappa}+\frac{R}{A}\right)\Psi.
\end{align}
Let us
define the Hilbert space as $\mathcal{H}=L^2(\mathbb{R}_{+},dA)$.
Then $P_A$ is
Hermitian as well as in the case of ordering (a). 
The condition for the Hamiltonian (\ref{ham2}) 
to be self-adjoint is also given by (\ref{BC}).

The dimensionless Schr\"{o}dinger equation is given by
\begin{align}
i\frac{\partial \Psi}{\partial \eta} = \frac{1}{x}\frac{\partial^2 \Psi}{\partial x^2}-\frac{1}{x^2}\frac{\partial\Psi}{\partial x}+\left(\frac{2}{3x^3}+ \lambda x^3 + \frac{r}{x}\right)\Psi.\label{Schrodinger2}
\end{align}

\subsection{Operator ordering (c) $\frac{P^2_{A}}{A} \to \frac{1}{\hat{A}}\hat{P}^2_A$ }
In the case (c), the Hamiltonian operator is given by
\begin{align}
\hat{H}=-  \frac{\kappa}{12} \frac{1}{\hat{A}}\hat{P}^{2}_A+ \frac{\Lambda \hat{A}^{3}}{\kappa}+\frac{R}{\hat{A}},
\end{align}
and the Schr\"{o}dinger equation is written in the simplest form
\begin{align}
i\hbar \frac{\partial \Psi}{\partial \tau} = \frac{\kappa \hbar^2}{12A}\frac{\partial^2 \Psi}{\partial A^2} + \left(\frac{\Lambda A^3}{\kappa}+\frac{R}{A}\right)\Psi.
\end{align}
Although we have defined the Hilbert space naturally as $\mathcal{H}=L^2(\mathbb{R}_{+},dA)$ in the previous two cases, here one cannot obtain specific boundary conditions with that definition.
In order to avoid the difficulty, we shall choose the Hilbert space as $\mathcal{H}=L^2(\mathbb{R}_{+},AdA)$ consisting of square integrable functions of $A$ with respect to the measure $AdA$.
With this inner product, $P_A$ cannot be Hermitian.
It can be shown that the boundary condition on the wave functions 
for $\hat H$ to be
self-adjoint is $\Psi
\bigr|_{A=0}=\gamma\frac{d\Psi}{dA}\bigr|_{A=0}$, 
which is discussed in the context of the Wheeler-DeWitt theory~\cite{Lemos,Lemos2}.
The two simplest cases with $\gamma=0$ and $\gamma=\infty$ are
\begin{align}
(1)\ \Psi\bigr|_{A=0}=0 \quad \textrm{and} \quad 
(2)\ \frac{d\Psi}{d A}\biggr|_{A=0}=0.\label{bc}
\end{align}

The dimensionless Schr\"{o}dinger equation is written as 
\begin{align}
i\frac{\partial \Psi}{\partial \eta} = \frac{1}{x}\frac{\partial^2 \Psi}{\partial x^2}+\left(\lambda x^3 + \frac{r}{x}\right)\Psi.\label{Schrodinger3}
\end{align} 
The boudary conditions (\ref{bc}) and Eq. (\ref{Schrodinger3}) will be discussed in Appendix A.

\section{Dynamics of the universe \label{sec6}}

\begin{figure}[htb]
 \begin{center}
  \subfigure[]{ 
   \includegraphics[width=.45\columnwidth]{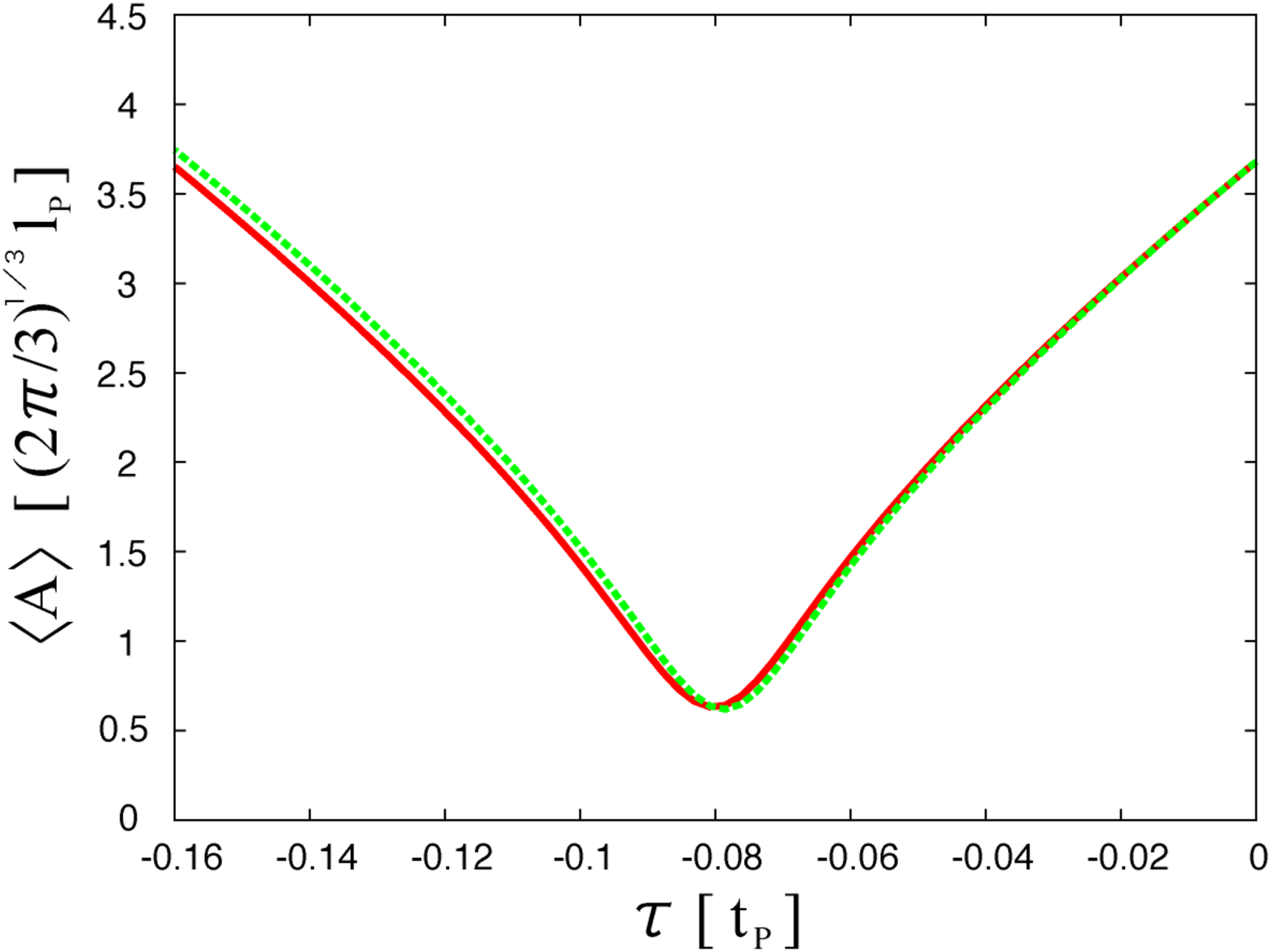}
  \label{fig:exp1}}\\ 
  \subfigure[]{
   \includegraphics[width=.45\columnwidth]{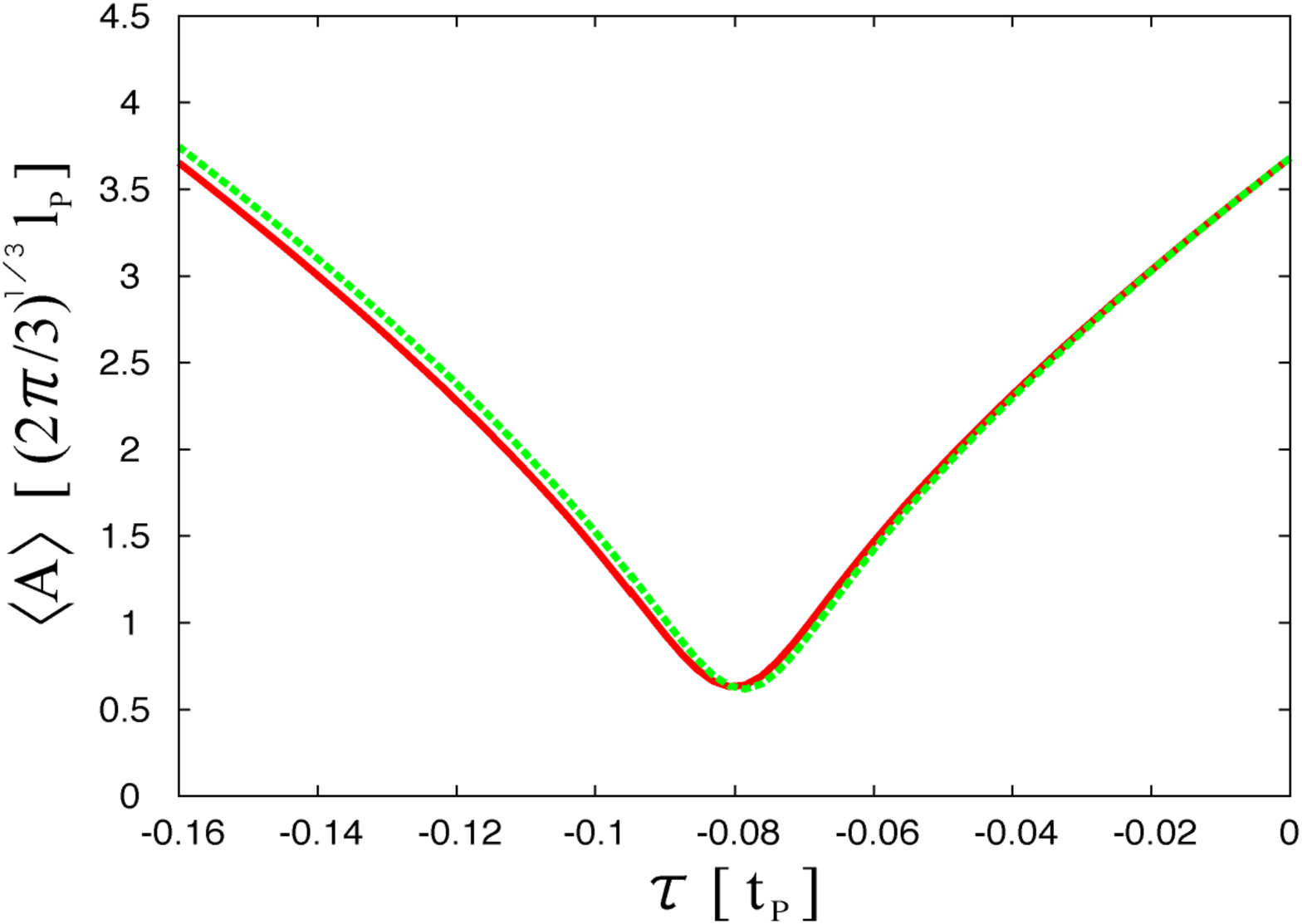}
  \label{fig:exp2}}~
  \subfigure[]{
   \includegraphics[width=.45\columnwidth]{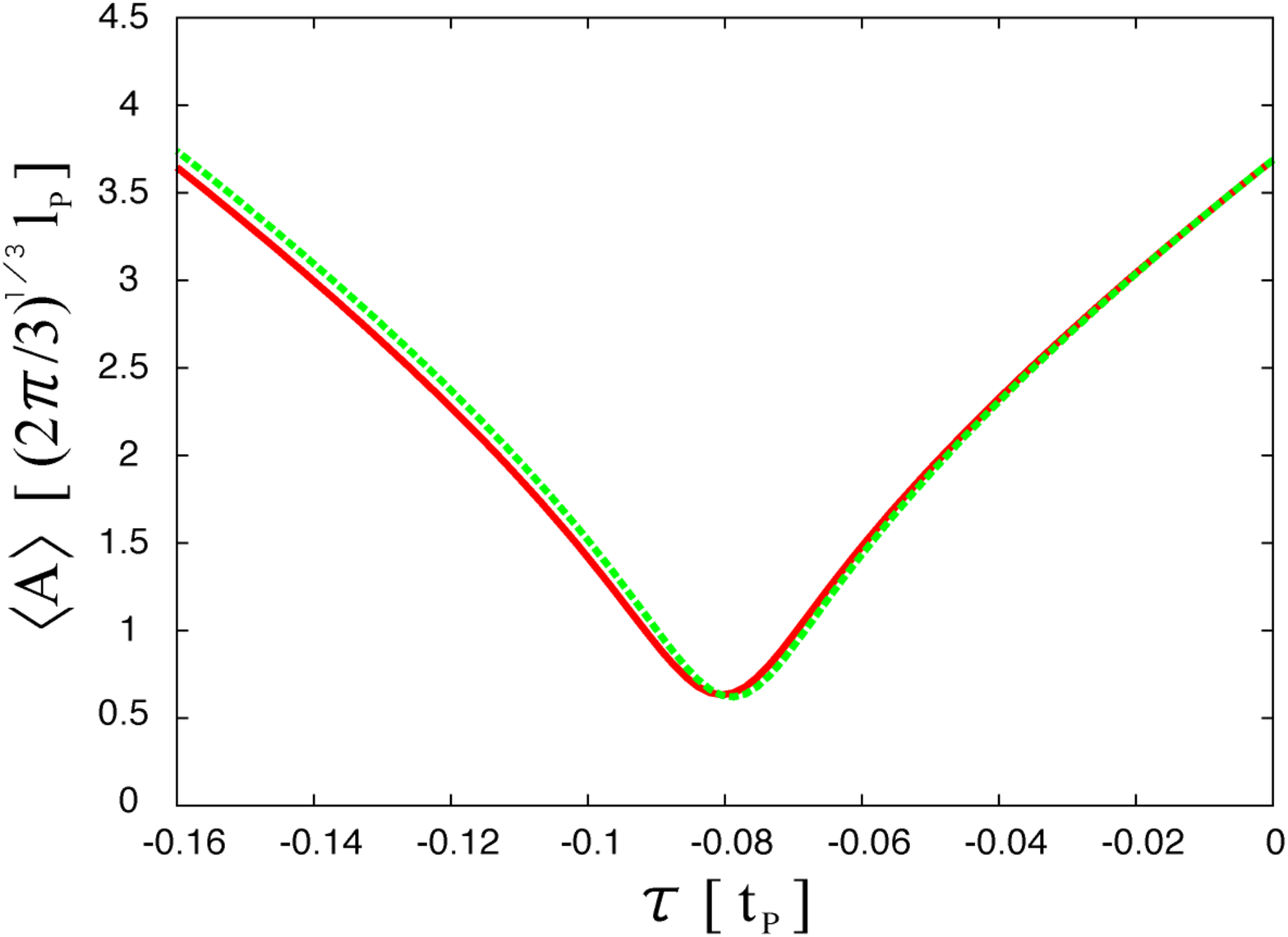}
  \label{fig:exp3}}
  \caption{The expectation value of the scale factor
  $\langle A\rangle$ as a function of time $\tau$ for the four cases $(r,\lambda)=(0,0),(0,1),(1,0)$, and $(1,1)$, where $r$ represents the dimensionless parameter associated to the total energy of the radiation while $\lambda$ is the dimensionless cosmological constant.
  We show only the cases when the matter involves the radiation because the plots in the models with or without the radiation almost completely overlap.  
  The solid line represents the cases with the cosmological constant, while the dashed line is for the cases without the cosmological constant.
  Figs.(a), (b), and (c) correspond to the three kinds of the operator ordering discussed in Sec.\ref{sec5}.} 
  \label{fig:exp}
 \end{center}
\end{figure}

\begin{figure}[htb]
 \begin{center}
  \subfigure[]{ 
   \includegraphics[width=.45\columnwidth]{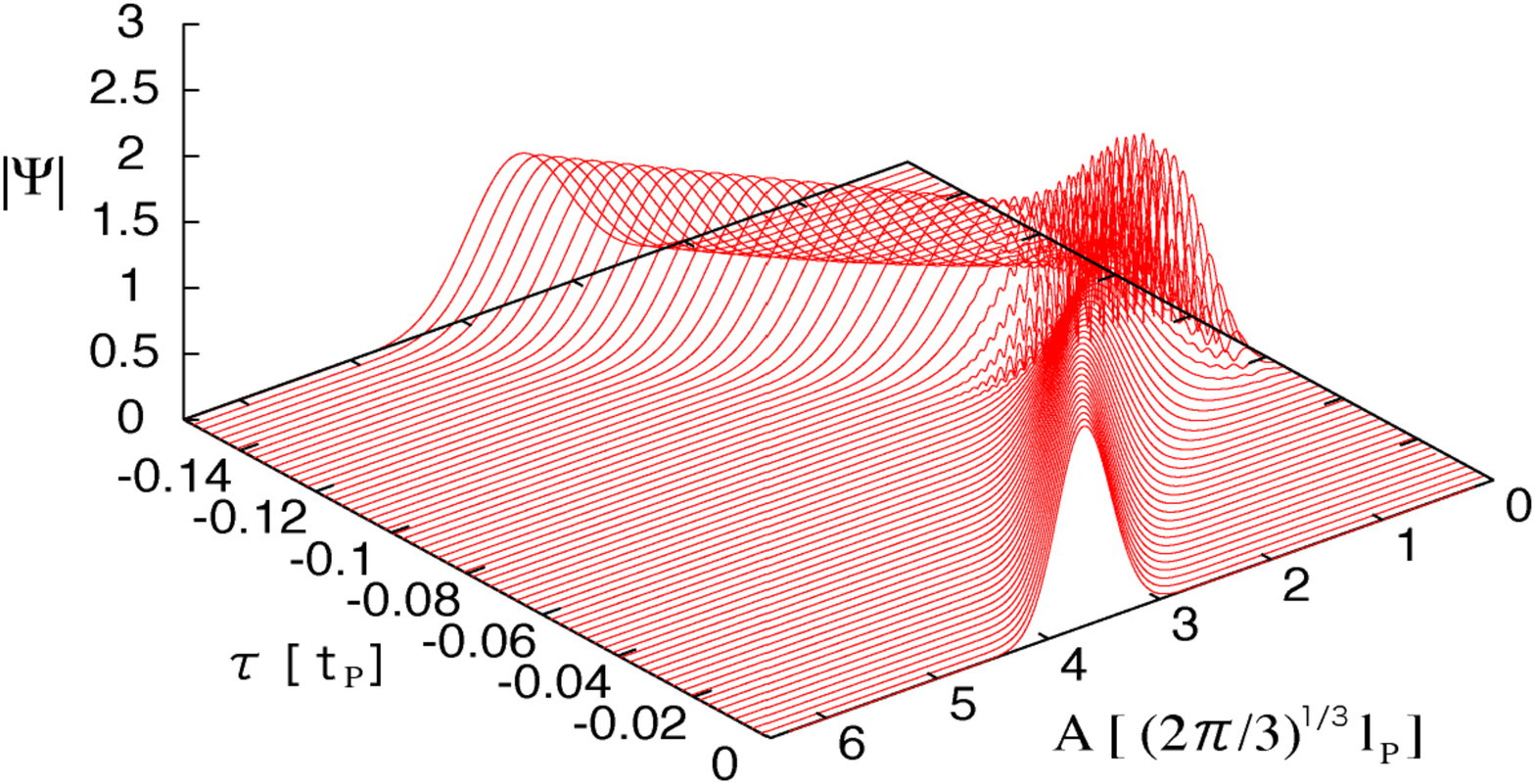}
  \label{fig:wave1}}\\ 
  \subfigure[]{
   \includegraphics[width=.45\columnwidth]{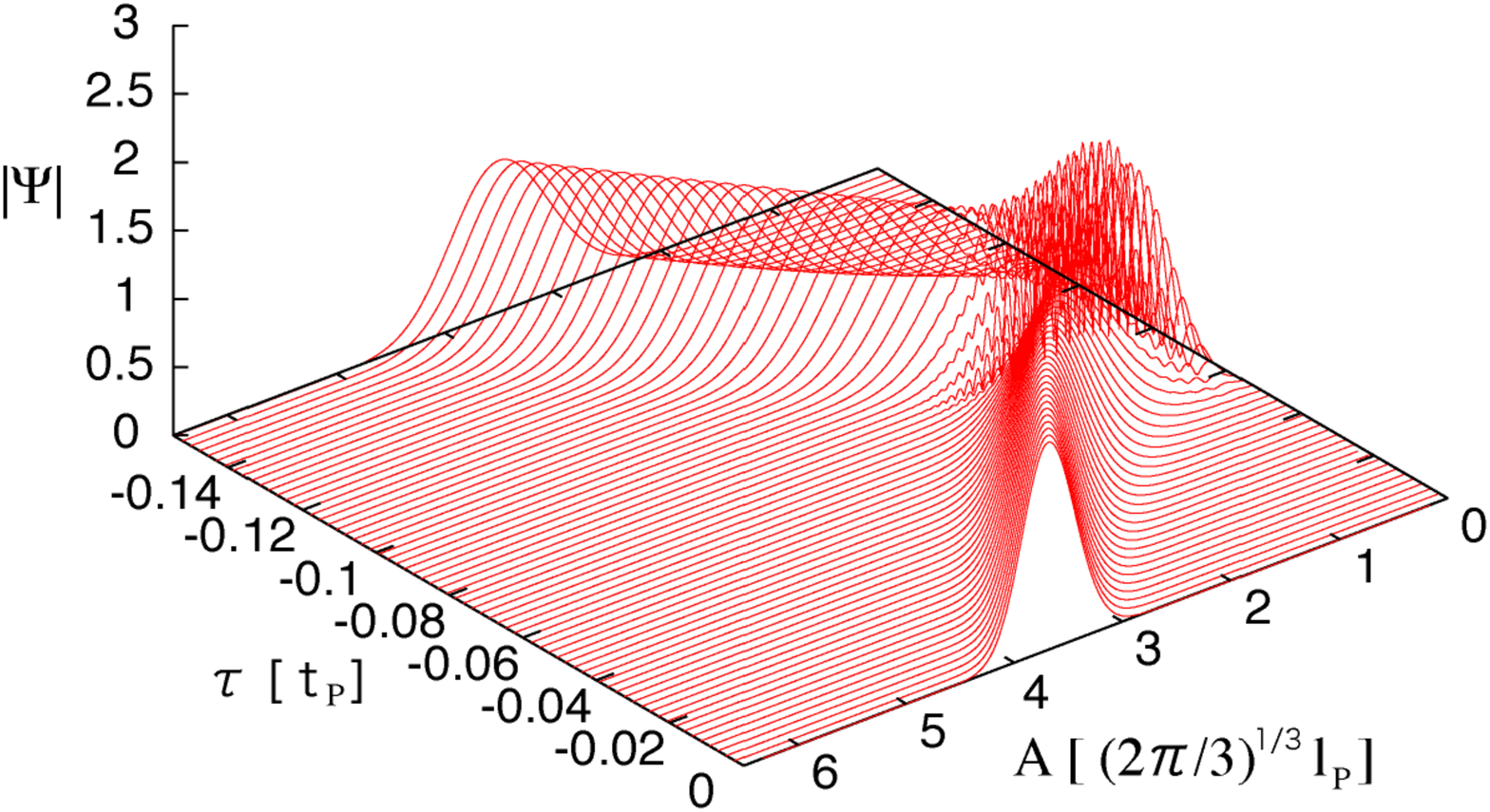}
  \label{fig:wave2}}~
  \subfigure[]{
   \includegraphics[width=.45\columnwidth]{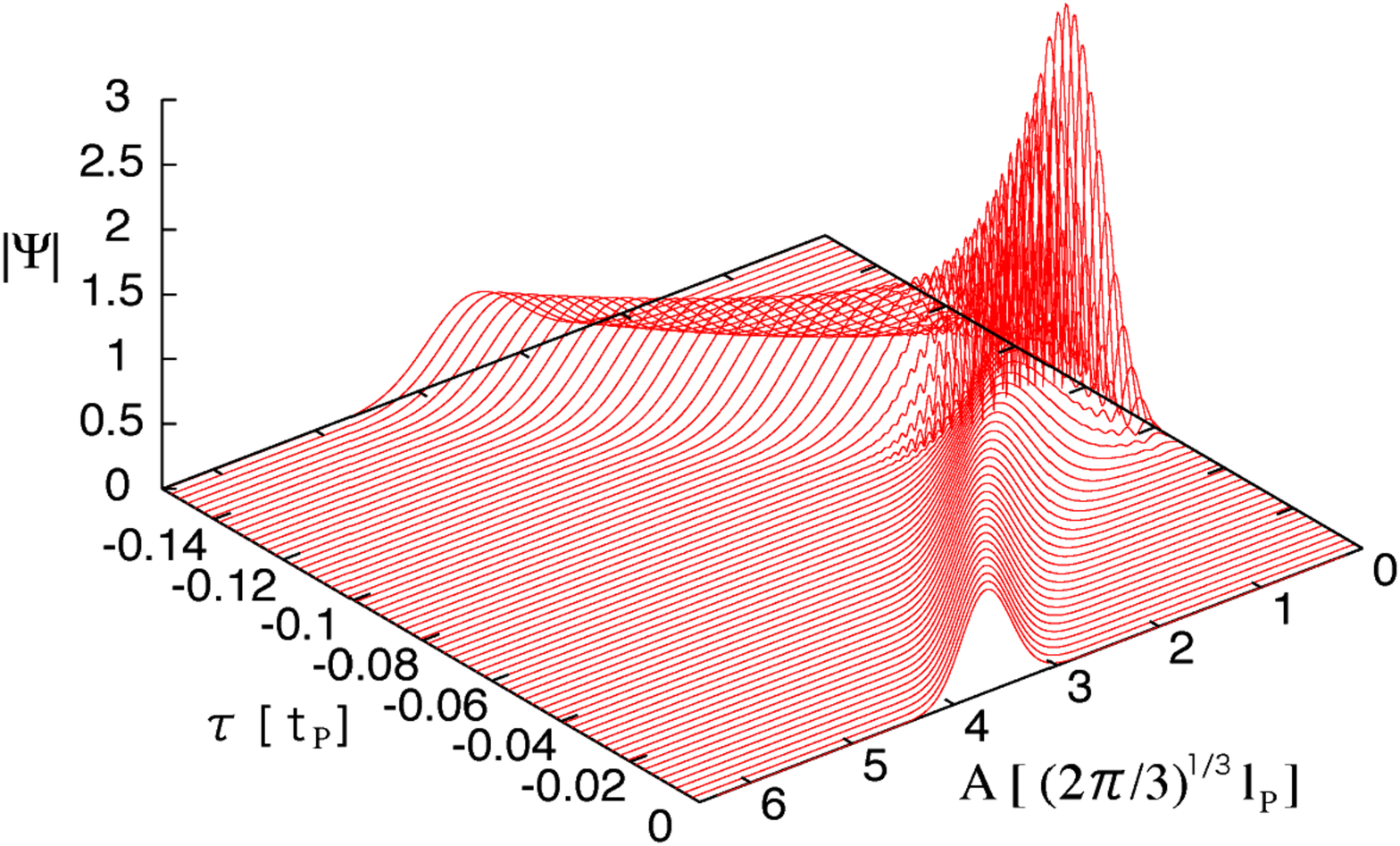}
  \label{fig:wave3}}
  \caption{The absolute value of the wave function is plotted as a function of the time $\tau$ and the scale factor $A$ in the case when the matter consists of the dust, the radiation, and the cosmological constant. Figs.(a), (b), and (c) correspond to the three kinds of the operator ordering discussed in Sec.\ref{sec5}.} 
  \label{fig:wave}
 \end{center}
\end{figure}

In this section, we shall analyze the dynamics of the Universe.
We first consider the case (a) discussed in the previous section.
We perform numerical calculations in order to solve the Schr\"{o}dinger equation (\ref{Schrodinger}) because one cannot analytically solve the equation.
The procedure of the numerical calculations is as follows. 
We first set the initial
wave function sharply peaked at some value of $A$. 
We next evolve it backward in time and evaluate the expectation value of the scale factor.
The boundary condition considered here is (\ref{bc2}).
The numerical methods used here are the fourth-order Runge-Kutta method in the time integration and the midpoint difference method for the spatial differentiation.
Although the equation to solve is (\ref{Schrodinger}), 
it is not simple to control the errors in the whole range of $x$ because 
of the rapidly growing potential term proportional to $x^3$. 
Therefore, 
we change the spatial variable to $y:=x^{\frac{3}{2}}$, 
so that the Schr\"{o}dinger equation 
 (\ref{Schrodinger}) becomes 
\begin{align}
i\frac{\partial \Psi(y,\eta)}{\partial \eta} = \frac{9}{4}\frac{\partial^2 \Psi(y,\eta)}{\partial y^2} &- \frac{3}{4}\frac{1}{y}\frac{\partial \Psi(y,\eta)}{\partial y} \nonumber\\
+ &\left(\lambda y^2 + ry^{-\frac{2}{3}}\right)\Psi(y,\eta).\label{y}
\end{align}
We choose the initial wave function as
\begin{align}
\Psi(y,0)\propto y\exp\left(-\frac{(y-y_0)^2}{4\sigma^2}-ik_0y\right),\label{init}
\end{align}
which satisfies the boundary condition (\ref{bc2}).
Although the other choices of the initial wave function are possible,
one can obtain qualitatively the same results irrespective of them. 
Fig. \ref{fig:exp1} shows the comparison between the expectation values of the scale factor for the four cases $(r,\lambda)=(0,0),(0,1),(1,0)$, and $(1,1)$, where we set the initial conditions as $y_0=7$, $k_0=20$, and $\sigma=0.5$.
The absolute values of the wave function when $r=\lambda=1$ are plotted in Fig. \ref{fig:wave1}.
The effect from the radiation is so small that the plots 
for the models with or without the radiation 
almost completely
overlap.
Therefore, we only show the two cases when the matter involves the radiation, with or without the cosmological constant, in Fig. \ref{fig:exp1}.   
As the figure indicates, the expectation value of the scale factor has a nonzero minimum, that is, the initial singularity is replaced by a big bounce. 
It also can be seen from Fig. \ref{fig:exp1} that the effect of a cosmological constant slightly decelerates the expansion of the Universe.
We note that the values of the cosmological constant and the energy of the
radiation used here are much larger than those suggested from the astrophysical
observations.
Therefore, the actual contributions from the cosmological constant and the radiation are still smaller, so that the results obtained here can be thought as the universal feature. 

We also perform the above calculations in cases (b) and (c).
The expectation value of the scale factor and the absolute value of the wave function in case (b) are illustrated in Figs. \ref{fig:exp2} and \ref{fig:wave2} respectively, and Figs. \ref{fig:exp3} and \ref{fig:wave3} are for case (c). 
As Fig. \ref{fig:exp} indicates, we obtain almost identical behavior, qualitatively and quantitatively, of the expectation value of the scale factor irrespective of the operator ordering.
We see from Fig. \ref{fig:wave} that the peak of the wave packet in case (c) is lower in comparison than with cases (a) and (b).

A special feature of case (c) is that one can analytically solve the Schr\"{o}dinger equation, construct wave packets and compute the expectation value of the scale factor when the matter consists only of the dust.
We show that in Appendix A and the same calculations appear in the context of the Wheeler-DeWitt theory~\cite{Lemos2}, where observables are not gauge-invariant.

\section{Conclusion \label{sec7}}
We have constructed and analyzed a gauge-invariant quantum theory of the FRW universe with dust.
In order to obtain the quantum theory free from the problem of time and observables, we have used the following two properties: (i) the Lagrangian with the dust introduced by Brown and Kucha\v{r} enables one to deparametrize the system, that is, the Hamiltonian constraint has the form $C=P+h(q^a,p_a)=0$ and (ii) deparametrized theories allow the construction of the reduced phase space coordinatized by gauge-invariant quantities through the relational formalism.
We have made a natural choice of the time variable which agrees with the cosmological proper time when one solves the equations of motion.
Then, we have quantized the reduced system and obtained the Schr\"{o}dinger equation.
In order to analyze whether the initial singularity is avoided by the quantum gravitational effects, we have constructed wave packets, evolved them backward in time and evaluated the expectation value of the scale factor.
It has been shown that the expectation value of the scale factor never goes to zero, that is, the initial singularity of the Universe is replaced by a big bounce.
We have obtained qualitatively and quantitatively almost identical results irrespective of three operator orderings. 
Although we have used a specific boundary condition in Sec.V, it is not necessarily responsible for the results.
Indeed, it is shown in Appendix A that a bouncing universe is obtained under two different boundary conditions.
The construction of a quantum theory discussed in this paper can be extended to other models, e.g., the FRW universe with a scalar field, and we shall study these problems in future works.

\appendix
\section{Dust universe in the case (c)}

In case (c), when the matter consists only of the dust, one can analytically solve the Schr\"{odinger} equation, construct wave packets and compute the expectation value of the scale factor. 
The boundary conditions considered here are (\ref{bc}). 
From the Schr\"{o}dinger equation (\ref{Schrodinger3}), the energy eigenvalue equation reads
\begin{align}
\frac{1}{x}\frac{d^2\psi(x)}{dx^2}=E\psi(x).\label{s}
\end{align}
The general solution of the equation (\ref{s}) is 
\begin{align}
\psi(z)=C_1 \textrm{Ai}(-z)+C_2 \textrm{Bi}(-z),\label{ai}
\end{align}
where $z:=(-E)^{\frac{1}{3}}x$, $\textrm{Ai}$ and $\textrm{Bi}$ are the Airy functions and $C_{j}$'s are arbitrary constants.
If $z$ has the sign opposite to $x$, that is, if $E\ge 0$, it is easy to see
from the behavior of the Airy functions that there are no wave functions which
has a finite norm and satisfy the boundary conditions (\ref{bc}).
Therefore, the energy eigenvalue $E$ must be restricted to the range $E< 0$.
This fact is reasonable because the total energy, which consists of the
gravitational energy and the matter energy, is constrained to vanish and $E$ can
be interpreted as the energy of gravity. 
Under the restriction $E<0$, we can transform the wave function (\ref{ai}) by
using the relation between the Airy functions and the Bessel functions
$J_{\pm\frac{1}{3}}$: 
\begin{align}
\psi(z)=\sqrt{z}\left[ C_1 J_{\frac{1}{3}} \left( \frac{2}{3}z^{\frac{3}{2}}\right)+C_2 J_{-\frac{1}{3}}\left( \frac{2}{3}z^{\frac{3}{2}}\right) \right].
\end{align}

It follows from the behavior of the Bessel functions for small argument that the allowed wave functions under each of the boundary conditions (\ref{bc}) are
\begin{align}
&(1)\ \psi_E^{(1)}(z)=\sqrt{z}J_{\frac{1}{3}} \left( \frac{2}{3}z^{\frac{3}{2}}\right), \nonumber\\
&(2)\ \psi_E^{(2)}(z)=\sqrt{z}J_{-\frac{1}{3}} \left( \frac{2}{3}z^{\frac{3}{2}}\right).
\end{align}  
Although these solutions are not normalizable with respect to the measure $AdA$, we can construct wave packets which have a finite norm by superposing them. 
This situation is similar to the case of a free particle in quantum mechanics where the energy eigenstates are not normalizable and one often constructs a wave packet to see the motion of the particle.
 
The general solutions of the Schr\"{o}dinger equation $i\frac{\partial \Psi}{\partial \eta}=\frac{1}{x}\frac{\partial^2 \Psi}{\partial x^2}$ are written in the form    
\begin{align}
\Psi^{(\textrm{I})}&(x,\eta)=\int^{0}_{-\infty}C^{(\textrm{I})}(E)e^{-iE\eta}\psi^{(\textrm{I})}_EdE\nonumber\\
&=\sqrt{x}\int^{\infty}_{0}C^{\prime (\textrm{I})}(\epsilon)e^{i\frac{9}{4}\epsilon^2 \eta}\epsilon^{\frac{1}{3}+1}J_{\pm \frac{1}{3}}\left(\epsilon x^{\frac{3}{2}}\right)d\epsilon,
\end{align}
where $\epsilon=\frac{2}{3}\sqrt{-E}$ and $\textrm{I}=1,2$.
If we choose the functions $C^{\prime (\textrm{I})}(\epsilon)$ to be $C^{\prime (1)}(\epsilon)=e^{-\alpha \epsilon^2}$ and $C^{\prime (2)}=\epsilon^{-\frac{2}{3}}e^{-\alpha \epsilon^2}$ where $\alpha$ is an arbitrary positive constant, we can use the formula \cite{formula} for the Bessel function $\int^{\infty}_{0}e^{-ax^2}x^{\nu +1}J_{\nu}(bx)dx=\frac{b^{\nu}}{(2a)^{\nu+1}}e^{-\frac{b^2}{4a}}, \textrm{Re}(a)>0, \textrm{Re}(\nu)>-1$.
Then, the wave functions take the form of the following wave packets:
\begin{align}
&\Psi^{(1)}(x,\eta)=\frac{x}{(2\beta)^{\frac{4}{3}}}e^{-\frac{x^3}{4\beta}},\nonumber\\
&\Psi^{(2)}(x,\eta)=\frac{1}{(2\beta)^{\frac{2}{3}}}e^{-\frac{x^3}{4\beta}},\label{wave packet}
\end{align}
where $\beta:=\alpha - i\frac{3\pi}{2}\eta$.
We interpret the wave packets (\ref{wave packet}) as the states of the Universe.

The expectation value of $x$ is given by
$\langle x \rangle^{(\textrm{I})} = \int^{\infty}_{0}|\Psi^{(\textrm{I})}|^2 x^2dx/ \int^{\infty}_{0} |\Psi^{(\textrm{I})}|^2 xdx$
and we can calculate this integration with respect to the wave packets:
\begin{align}
&\langle x \rangle^{(1)}=\frac{2\Gamma \left(\frac{5}{3}\right)}{3\Gamma\left(\frac{7}{3}\right)}\left( \frac{\alpha^2 + \left( \frac{9}{16}\eta \right)^2}{2\alpha} \right)^{\frac{1}{3}},\nonumber\\
&\langle x \rangle^{(2)}=\frac{4}{3\cdot 2^{\frac{1}{3}} \Gamma\left(\frac{5}{3}\right)}\left( \frac{\alpha^2 + \left( \frac{9}{16}\eta \right)^2}{2\alpha} \right)^{\frac{1}{3}}.
\end{align}
These solutions show that the expectation value of the scale factor never goes to zero, that is, the initial singularity of the Universe is avoided by quantum gravitational effects.
The asymptotic behavior of the expectation values for large $x$ becomes
\begin{align}
\langle x \rangle^{(\textrm{I})}\propto \eta^{\frac{2}{3}}.
\end{align}
Thus, they are in good agreement with the classical trajectories when the Universe is sufficiently large.


\end{document}